\documentclass[10pt,twoside,a4paper]{article}
\input axodraw.sty
\def\lsim{\:\raisebox{-0.5ex}{$\stackrel{\textstyle<}{\sim}$}\:}

\usepackage{graphicx}
\begin{document}
\thispagestyle{empty} 
\title{
\vskip-2cm
{\baselineskip14pt
\centerline{\normalsize MZ-TH/99--39 \hfill} 
\centerline{\normalsize hep--ph/9909421\hfill}} 
\vskip1.5cm
Beyond the Standard Model at HERA\footnote{Lecture notes to appear in
  the proceedings of the Ringberg Workshop {\it New Trends in HERA
    Physics 1999}, May 30--June 4, 1999.} \\[3ex]
\author{H.~Spiesberger \\[2ex]
\normalsize{Institut f\"ur Physik,
  Johannes-Gutenberg-Universit\"at,}\\ 
\normalsize{Staudinger Weg 7, D-55099 Mainz, Germany} \\[2ex]
} }
\date{}
\maketitle
\begin{abstract}
\medskip
\noindent
  The prospects of physics beyond the standard model in deep inelastic
  scattering are reviewed, emphasizing the search for contact
  interactions, for leptoquarks and for supersymmetry with $R$-parity
  violation. $R$-parity violating supersymmetry is explored as a
  speculative source of events with high energy muons and missing
  transverse momentum, but no convincing explanation for events of this
  type observed at H1 is found.
\end{abstract}


\section{Introduction}

The luminosity delivered to the experiments at HERA has now become large
enough to open a new focus of physics analyses looking at processes with
cross sections of the order of 1\,pb and below. This is the typical
value for neutral and charged current (NC and CC) cross sections at
large values of Bjorken $x$ and momentum transfer $Q^2$.  Also
measurements of rare standard model (SM) processes like the production
of an additional gauge boson, are becoming possible.  These low cross
section processes provide a wealth of possibilities to look for
deviations from the standard model predictions and constitute important
backgrounds for searches for physics beyond the standard model
\cite{future}.

\begin{figure}[t]
\unitlength 1mm
\begin{minipage}[b]{0.5\textwidth}
\begin{picture}(57,65)
\put(0,0){
\includegraphics[width=\textwidth]{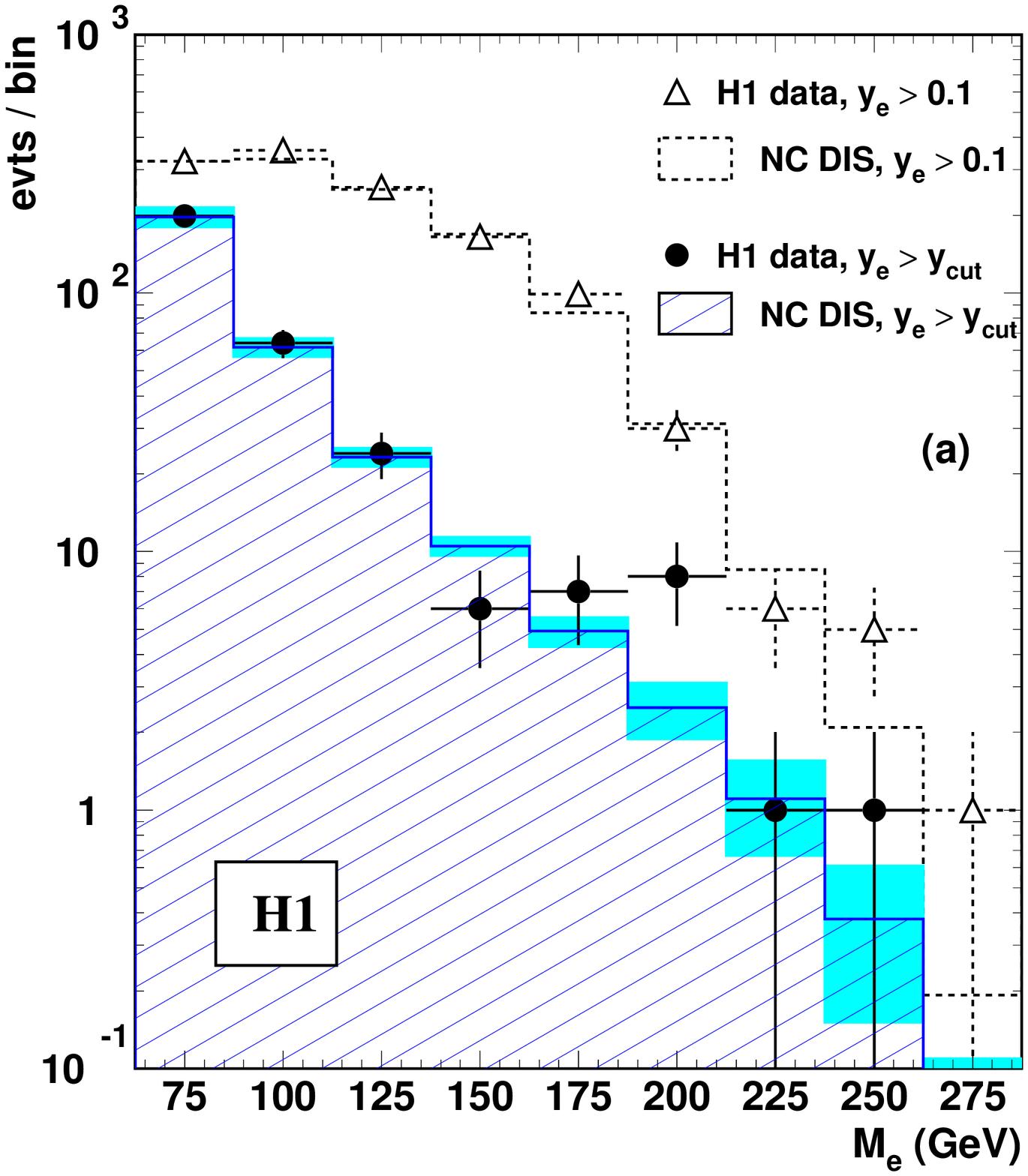}}
\end{picture}\par
\end{minipage}\hfill
\begin{minipage}[b]{0.5\textwidth}
\begin{picture}(57,65)
\put(0,0){
\includegraphics[width=\textwidth]{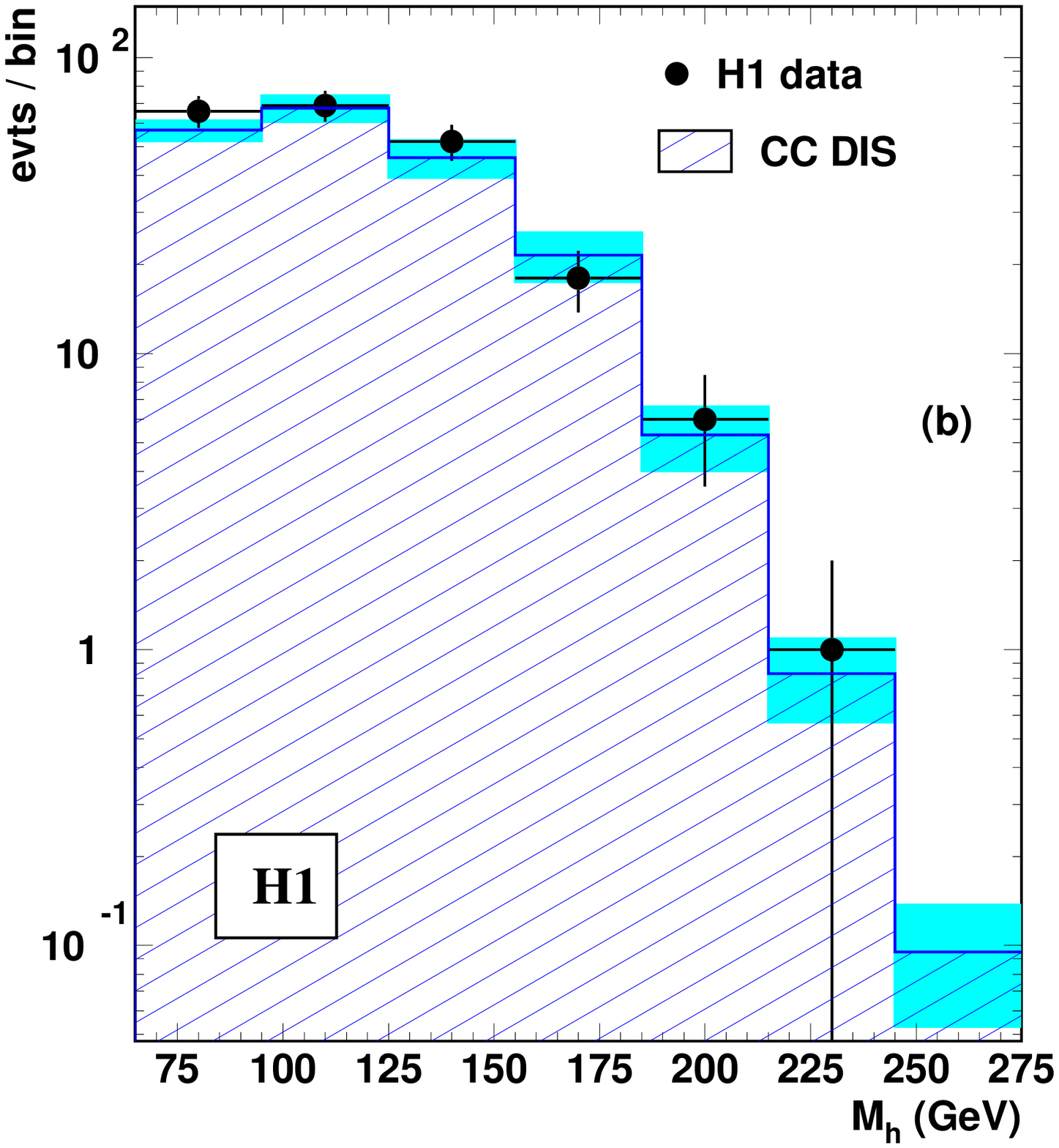}}
\end{picture}
\end{minipage}
\caption[]{Mass spectra for NC (left) and CC (right) DIS-like events for 
  data (symbols) and SM expectation (histograms) observed at H1 in
  37\,pb$^{-1}$ of $e^+p$ data taken in 1994--1997 \cite{h1lq}}  
\label{fig1}
\end{figure}

The motivation to search for new physics at HERA has received a strong
impetus by the observation of enhancements of cross sections at several
places. The excess of events at large $x$ and large $Q^2$ in NC and CC
scattering \cite{excess} observed in the 1994--96 $e^+p$ data has been
discussed at length in the literature (see \cite{altarelli,hsrr} and
references therein). A similar excess was not observed in the 1997 data
so that the significance in the complete 1994--97 data sample is
reduced, but still there: in the mass bin 200\,GeV$\pm \Delta M/2$
($\Delta M = 25$\,GeV) and for $y > 0.4$, H1 observed 8 events, but only
$2.87 \pm 0.48$ are expected (see Fig.\ \ref{fig1}a). In the CC channel
the observed number of events is in agreement with the SM predictions
within the uncertainties: H1 observed 7 events with $Q^2 >
15,000$\,GeV$^2$ ($4.8 \pm 1.4$ expected) and ZEUS found 2 with $Q^2 >
35,000$\,GeV$^2$ ($0.29 \pm 0.02$ expected), both in the 1994--96 data
set. Notably the occurrence of five events with an isolated muon and
large missing transverse momentum at H1 \cite{h1muon} which are
seemingly not all a sign of $W$ production presents a challenge for the
understanding of the experiments.

In the following, I selected some of the alternatives to standard model
physics which, if realized in nature, have a good chance to be
discovered at HERA. If not, HERA is expected to significantly contribute
to setting limits on their respective model parameters.  Other related
topics of interest have been discussed previously in Refs.\ 
\cite{future,durham}.


\section{New Physics Scenarios}

Despite of the great success of the standard model, various conceptual
problems provide a strong motivation to look for extensions and
alternatives. Two main classes of frameworks can be identified among the
many new physics scenarios discussed in the literature:
\begin{itemize}
\item Parametrizations of more general interaction terms in the
  Lagrangian like contact interactions or anomalous couplings of gauge
  bosons are helpful in order to {\it quantify the agreement} of
  standard model predictions with experimental results. In the event
  that deviations are observed, they provide a framework allowing to
  relate different experiments and cross-check possible theoretical
  interpretations. Being insufficient by themselves, e.g.\ because they
  are not renormalizable, parametrizations are expected to show the 
  directions to the correct underlying theory if deviations are
  observed. 
\item Models, sometimes even complete theories, provide specific
  frameworks that allow a consistent derivation of cross sections for
  conventional and new processes. Examples are the two-Higgs-doublet
  extension of the standard model, grand unified theories and, most
  importantly, supersymmetry with or without $R$-parity violation.
\end{itemize}
The following examples attained most interest when the excess of
large-$Q^2$ events at HERA was made public \cite{altarelli,hsrr}. 


\subsection{Contact interactions}

The contact interaction (CI) scenario relevant for NC processes assumes
that 4-fermion processes are modified by additional terms in the
interaction Lagrangian of the form
\begin{equation}
{\cal L}_{\rm CI} = 
  \sum_{\begin{array}{c}i,k = L,R\\ q=u,d,\cdots
  \end{array}} \eta^q_{ik} 
  \frac{4\pi}{\left(\Lambda^q_{ik}\right)^2} \left(\bar{e}_i
    \gamma^{\mu} e_i\right) \left(\bar{q}_k \gamma_{\mu} q_k\right) \; .
\label{CI}
\end{equation}

\begin{figure}[hp]
\small
\begin{picture}(160,130)(-80,0)
\ArrowLine(50,120)(96,74)
\ArrowLine(104,66)(150,20)
\ArrowLine(50,20)(96,66)
\ArrowLine(104,74)(150,120)
\thicklines
\Line(96,66)(104,66)
\Line(104,66)(104,74)
\Line(104,74)(96,74)
\Line(96,74)(96,66)
\put(39,122){$e_i$}
\put(155,122){$\bar{e}_i$}
\put(39,18){$q_k$}
\put(155,18){$\bar{q}_k$}
\put(8,72){HERA}
\put(8,62){$ep \rightarrow eX$}
\put(54,67){$\Rightarrow$}
\put(90,130){LEP}
\put(69,120){$e^+e^- \rightarrow {\rm hadr}$}
\put(96.5,107){$\Downarrow$}
\put(80,12){Tevatron}
\put(70,2){$p\bar{p} \rightarrow \ell^+\ell^- X$}
\put(96.5,27){$\Uparrow$}
\end{picture}
\caption[]{Schematic view of a contact interaction term.}
\label{fig2}
\end{figure}
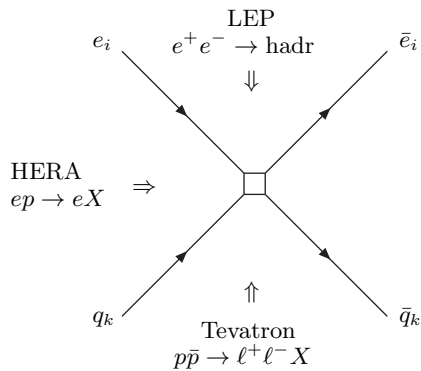
\noindent
Similar terms with 4-quark interactions would be relevant for new
physics searches at the Tevatron and 4-lepton terms would affect purely
leptonic interactions\footnote{Contact interactions modifying CC
  processes can be constructed in a similar way and have been
  investigated in Refs.\ \cite{agm,cornet}.}. In equation (\ref{CI}), as
is usual practice, only products of vector or axial-vector currents are
taken into account since limits on scalar or tensor interactions are
very stringent. Such terms are motivated in many extensions of the
standard model as effective interactions after having integrated out new
physics degrees of freedom like heavy gauge bosons, leptoquarks and
others, with masses beyond the production threshold. The normalization
with the factor $4\pi$ is reminiscent of models which predict CI terms
emerging from strong interactions at a large mass scale $\Lambda$.

Equation \ref{CI} predicts modifications of cross sections for processes
involving two leptons and two quarks in all channels as visualized in
Fig.\ \ref{fig2}.  Both enhancement or suppression are expected,
depending on the helicity structure of the contact term and its sign
$\eta^q_{ik}$. If the CI mass scale is large, the highest sensitivity is
expected at experiments with highest energies, but due to the extremely
high experimental precision, also atomic parity violation experiments at
low energies are sensitive to parity-odd combinations of helicities
\cite{apvlimits}.

Limits from single experiments at the Tevatron, HERA or LEP2 for models
with one single parameter \cite{cilimits} are typically in the order of
several TeV and all present high-energy experiments have achieved limits
in a very similar mass range despite of their different center-of-mass
energies. Consequently, with a signal at HERA one should expect visible
effects at LEP2 and at the Tevatron. Moreover, global fits taking into
account experimental data from these different sources give valuable
additional insight. Recent global fits \cite{zarnecki,cheung} have taken
into account new data from HERA, LEP2, Tevatron and CCFR. The resulting
limits for single-parameter models increase from the range 1.8--10.5 TeV
(derived from a single experiment) to 5.1--18.2 TeV (derived from the
global analysis) \cite{zarnecki}. In a general model where 8 independent
parameters are allowed to be non-zero at the same time, the limits are
of course weaker and range from 2.1 to 5.1 TeV for the various mass
scales $\Lambda_{ik}^q$. A comparison of various data obtained at LEP2,
Tevatron and HERA with the prediction of a model with contact
interactions as obtained in the best global fit of Ref.\ \cite{zarnecki}
shows that only the HERA data at highest values of $Q^2$ tend to support
the presence of a contact term.

\begin{figure}[hb]
\begin{center}
\includegraphics[width=0.8\textwidth]{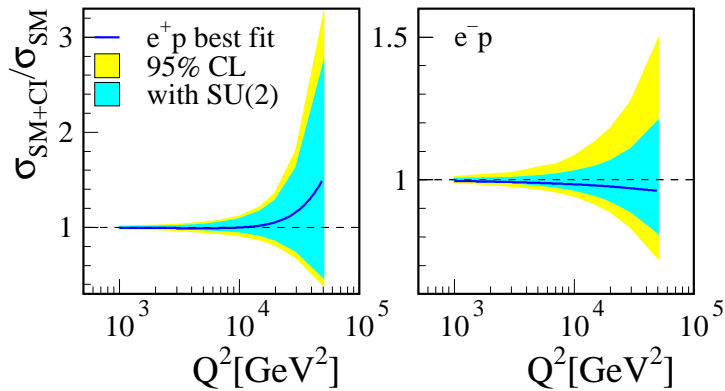}
\end{center}
\caption[]{The 95\,\%\,CL limit band on the ratios of $e^+$ and $e^-$
  cross sections for NC DIS at HERA with and without a contact term of
  the best fit of \cite{zarnecki2}}
\label{fig4}
\end{figure}

Assuming the presence of contact interactions with a mass scale in the
range allowed by the best fit one can derive 95\,\%\,CL limits for the
predicted deviations from the SM cross sections.  Figure \ref{fig4}
shows the results for $e^+p$ and $e^-p$ scattering at HERA. Obviously, a
possible deviation in electron scattering is much more restricted than
for positron scattering; in the latter case, deviations of the cross
section for $Q^2 > 15,000$\,GeV$^2$ from the standard model of $40\,\%$
are allowed, whereas only $20\,\%$ deviations are inside the 95\,\%\,CL
band for the former case. A luminosity of 100--200\,pb$^{-1}$ would
suffice in $e^+p$ scattering to observe such a deviation. On the other
hand, measurements with positrons at HERA have a better chance to
further improve limits on contact terms.

\begin{figure}[hptb]
\unitlength 1mm
\begin{minipage}[b]{0.5\textwidth}
\begin{picture}(57,66)
\put(0,0){
\includegraphics[width=\textwidth,bb=40 80 545 700]{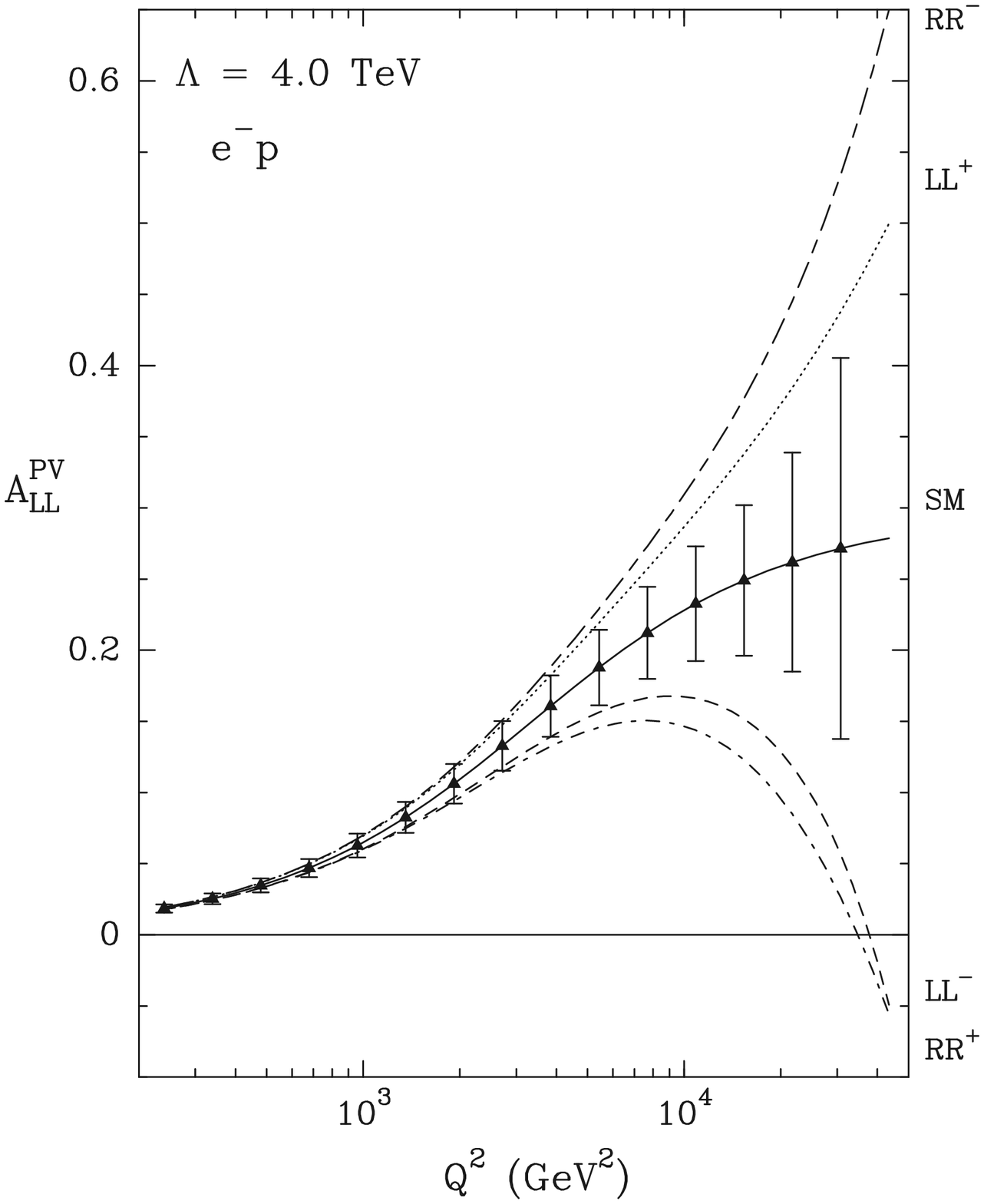}}
\end{picture}\par
\end{minipage}\hfill
\begin{minipage}[b]{0.5\textwidth}
\begin{picture}(68,75)
\put(0,0){
\includegraphics[width=\textwidth,bb=40 80 545 700]{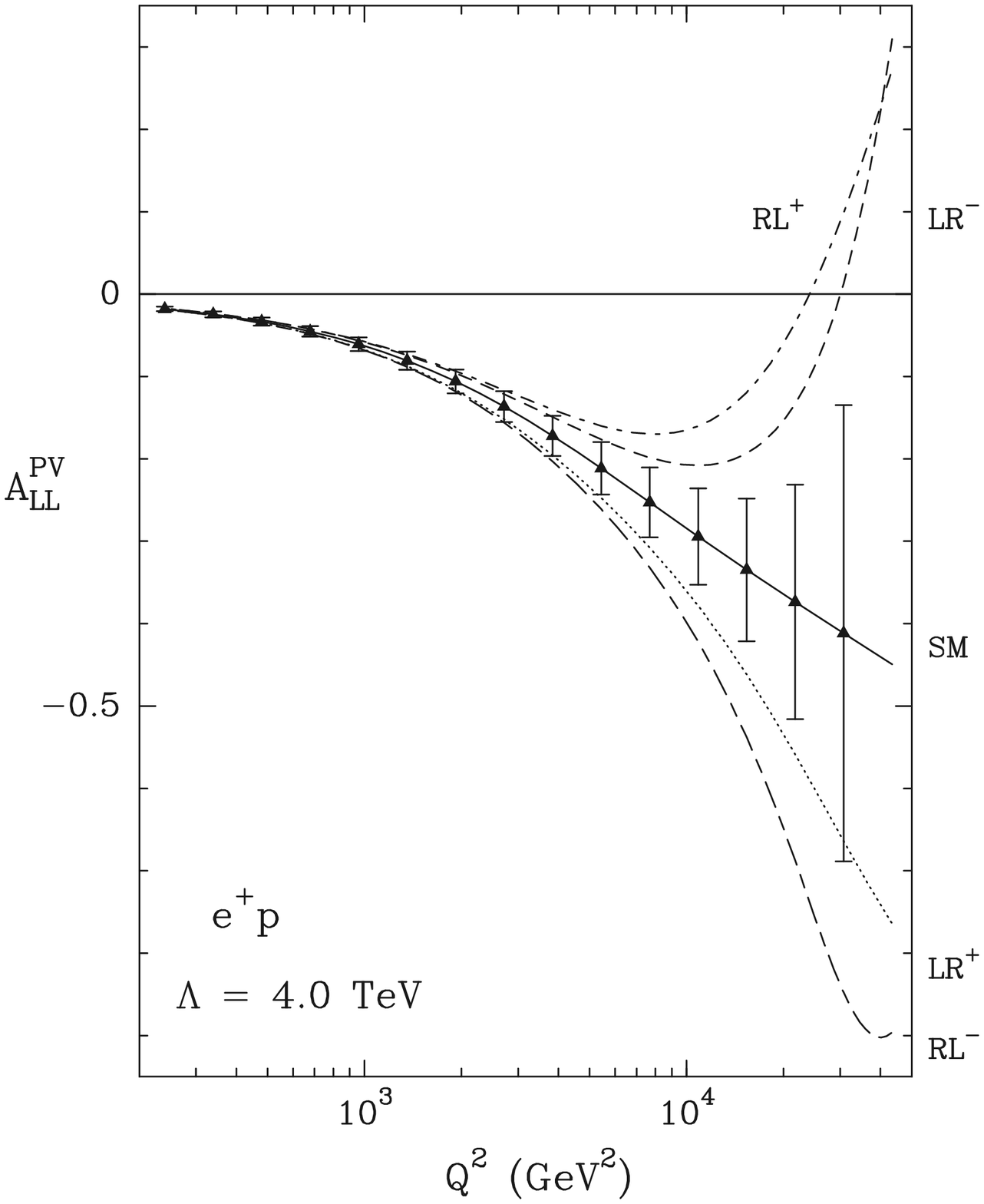}}
\end{picture}
\end{minipage}
\caption[]{Spin asymmetries $A_{LL}^{PV} (e^-)$ (left) and $A_{LL}^{PV}
  (e^+)$ (right). Solid lines correspond to the SM predictions; the
  expected errors are shown assuming a luminosity of 125\,pb$^{-1}$ for
  each configuration of beam polarizations. Non-solid lines correspond
  to CI scenarios with $\Lambda = 4$ TeV and helicities as indicated
  \cite{CIpol}}  
\label{fig5}
\end{figure}

In the case of the observation of deviations from the standard model
predictions, the combination of results obtained in different
experiments and from measurements with polarized beams will be helpful
to identify the helicity structure of contact interaction terms
\cite{CIpol}. This is visualized in Fig.\ \ref{fig5} where the
parity-violating spin-spin asymmetries in $e^{\pm}p$ scattering are
shown for models with various types of contact terms and a mass scale of
$\Lambda = 4$ TeV.  With an integrated luminosity of 125\,pb$^{-1}$ for
each configuration of beam polarization, these models would be clearly
distinguishable.


\subsection{Large Extra Dimensions}

In the usual contact term scenario, one concentrates on interaction
terms with mass dimension 6. Higher-dimension interactions are usually
assumed to be less important since they would be suppressed by a higher
power of the ratio of the center-of-mass energy and the mass scale
characterizing these interactions. Nevertheless it is interesting to
study the effect of such terms since models might exist where
higher-dimensional interactions are the dominating deviation from the
standard model framework. Recently, theories with large extra dimensions
emerging in low-scale compactified string theories have been shown to
constitute a viable alternative to the standard model \cite{LED}. A
specific class of these theories would predict deviations from standard
model cross sections through the exchange of gravitons and their
Kaluza-Klein excitations \cite{LED_ph}.  The effect can be described
with the help of dimension 8 NC contact terms \cite{LED_CI}, but there
would also exist completely new kind of interactions like electron-gluon
contact terms.

\begin{figure}[t]
\begin{center}
\includegraphics[width=\textwidth]{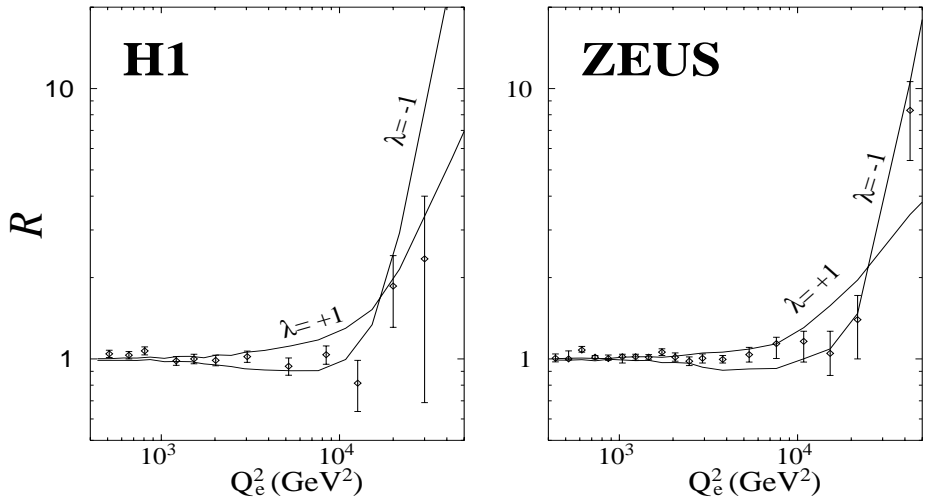}
\end{center}
\caption[]{Illustration of the effect of large extra dimensions on NC
  $e^+p$ scattering at HERA \cite{raychau}}
\label{fig6}
\end{figure}

Figure \ref{fig6} shows an example \cite{raychau} for the effect of
graviton exchange with two choices for the relative sign of the standard
model and new physics amplitudes compared to the large-$Q^2$ data from
H1 and ZEUS. The mass scale $M_s$ of such theories are chosen in this
example to saturate the 95\,\%\,CL limits: 543 (567)\,GeV for H1 (ZEUS)
data and $\lambda = +1$ and 436 (485)\,GeV for $\lambda = -1$. As
discussed in \cite{rizzo}, with an integrated luminosity of
250\,pb$^{-1}$ for each of electron and positron scattering with left-
and right-handed longitudinal polarization (i.e.\ 1\,fb$^{-1}$ in
total), HERA could set limits slightly above 1 TeV and would thus be
competitive with LEP2 (expected 1.1 TeV 95\,\%\,CL limit), but slightly
worse than the Tevatron (1.3 TeV). A future $e^+e^-$ linear collider
would be sensitive to mass scales above 4 TeV and the LHC can be
expected to shift the corresponding limit to 6.0 TeV \cite{rizzo}.


\subsection{Leptoquarks}

Leptoquarks appear in extensions of the standard model involving
unification, technicolor, compositeness, or $R$-parity violating
supersymmetry. In addition to their couplings to the standard model
gauge bosons, leptoquarks have Yukawa-type couplings to lepton-quark
pairs which allow their resonant production in $ep$ scattering.  The
generally adopted BRW-framework \cite{BRW} is based on only a few
assumptions concerning these Yukawa interactions which lead to a rather
restricted set of allowed states and the branching fractions $\beta_e$
for their decay to a charged lepton final state can only be 1, 0.5, or
0.  States which can be produced in $e^+u$ or $e^+d$ scattering have
$\beta_e = 1$ and for masses below 242\,GeV they are excluded by
Tevatron bounds.

Renewed theoretical work on the phenomenology of leptoquarks (see
\cite{hsrr} and references therein) was initiated by the observation of
an excess of events at large $x$ and large $Q^2$ in the 1994--96 HERA
$e^+p$ data which showed that the BRW-framework may indeed be too
restrictive. The crucial and least well motivated assumption there is
that leptoquarks are not allowed to have other interactions besides
their gauge and Yukawa couplings. In fact, most concrete models with
leptoquarks do predict additional interactions which may lead to decay
modes to other than lepton-quark final states. This would be interesting
since for example the Tevatron bounds do not exclude leptoquarks with
masses above 200\,GeV in scenarios with branching ratios $\beta_e \lsim
0.7$ \cite{altarelli,hr}.

A few examples for more general scenarios have been discussed in detail
in the literature. In Ref.\ \cite{babu} a model was proposed where two
leptoquark states show mixing induced by coupling them to the standard
model Higgs boson.  Alternatively, interactions to new heavy fields
might exist which, after integrating them out, could lead to leptoquark
Yukawa couplings as an effective interaction \cite{agm}, bypassing this
way renormalizability as a condition since this is assumed to be
restored at higher energies.  In the more systematic study of Ref.\ 
\cite{hr}, LQ couplings arise from mixing of standard model fermions
with new heavy fermions with vector-like couplings and taking into
account a coupling to the standard model Higgs. The most interesting
extension of the generic leptoquark scenario is, however,
$R_p$-violating supersymmetry which is discussed in the next subsection.

\begin{figure}[bh]
\begin{center}
\includegraphics[height=0.45\textheight,bb=90 330 280 580]{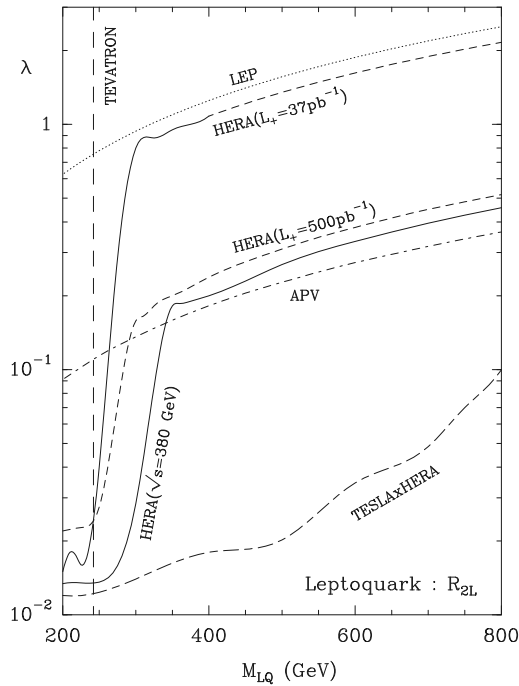}
\end{center}
\caption[]{Discovery limits for a scalar leptoquark at various collider
  experiments \cite{virey}}
\label{fig9}
\end{figure}
Searches at HERA \cite{kuze} are essential to exclude such more general
scenarios.  Despite of the strong dependence on the lepton-quark-LQ
Yukawa coupling $\lambda$, exclusion limits from HERA experiments cover
much larger mass values for small $\lambda$ than those obtained from
indirect searches at LEP2. Since the dependence on the branching ratio
in more general scenarios can be reduced considerably by combining NC
and CC data, HERA limits also supersede those from Tevatron for small
$\beta_e$. The most recent published limits from H1 \cite{h1lq} take
into account the finite width of LQ states and the interference of their
production amplitude with the standard model background, both effects
which turned out to be non-negligible for very large LQ masses. Scalar
leptoquarks with masses up to 275\,GeV and vector states up to 284\,GeV
are excluded at 95\,\%\,CL for $\lambda = e$ \cite{h1lq}. Similar mass
exclusion regions have been reported by ZEUS at recent conferences
\cite{zeuslq}. As shown in Fig.\ \ref{fig9} \cite{virey}, exclusion
limits for the coupling $\lambda$ at the same LQ mass values can
expected to be reduced by a factor of $\sim 5$ with 500\,pb$^{-1}$ of
$e^+p$ data. With this luminosity, limits on $\lambda$ for much larger
LQ masses from HERA will also come close to the limits following from
atomic parity violation experiments via the corresponding induced
contact interactions.  To further extend the search to large LQ-masses
and small Yukawa couplings, an increase of the center-of-mass energy of
$ep$ collisions (for example like at TESLA$\times$HERA) would be
essential.


\subsection{$R_p$-violating supersymmetry}

The Lagrangian of a supersymmetric version of the standard model may
contain a superpotential of the form
\begin{equation}
\begin{array}{lll}
W_{\not R_p} = & \phantom{+} \lambda_{ijk} L_i L_j E_k^c & 
  \hspace{10mm} \not\!\! L \\[1mm] 
& + \lambda'_{ijk} L_i Q_j D_k^c & 
  \hspace{10mm} \not\!\! L ~~~{\rm
    (includes~LQ}\raisebox{0.5ex}[-0.5ex]{\rule{1.1mm}{0.25mm}}{\rm
    like~couplings)} 
\\[1mm]
& + \lambda''_{ijk} U_i^c D_j^c D_k^c & \hspace{10mm} \not\!\! B
\end{array}
\label{rpviol}
\end{equation}
$L_i$ and $Q_i$ are the superfields for lepton and quark doublets and
$E_i^c$, $U_i^c$, $D_i^c$ the corresponding charge-conjugated ones for
charged leptons, up and down quarks, respectively and $i$, $j$, $k$ are
generation indices. The separate contributions in $W_{\not R_p}$ violate
lepton or baryon number conservation as indicated.  Imposing symmetry
under $R$-parity (defined as $R_p = (-1)^{3B+L+2S}$, $=1$ for particles
and $=-1$ for their superpartners) forbids the presence of $W_{\not
  R_p}$.  The phenomenology of supersymmetry with $R_p$ symmetry has
been searched for at all present high energy experiments and HERA may
set interesting limits which are complementary to those obtained at the
Tevatron \cite{future}.

Many low- and high-energy experiments put limits on the couplings
contained in $W_{\not R_p}$ \cite{add}; however, they do not forbid
interactions of the form $L_i Q_j D_k^c$ proportional to
$\lambda'_{ijk}$ in general, provided the $\lambda''_{ijk}$ are chosen
to be zero at the same time. This makes squarks appear as leptoquarks
which can be produced on resonance in lepton-quark scattering.  In
contrast to the generic leptoquark scenarios described above, squarks do
not only decay into lepton-quark final states via their $R_p$-violating
interactions but they can also decay into final states involving gauge
bosons or gauginos. These $R_p$-conserving decays lead to a large number
of interesting and distinct signatures (see \cite{dreiner} and
references therein)\footnote{Monte Carlo tools needed in searches for
  $R_p$-violating supersymmetry at HERA have been improved recently
  \cite{heramc}.}.  Characteristically one expects multi-lepton and
multi-jet final states. Mass and coupling parameters of $R_p$-violating
supersymmetry can be varied such that the branching ratio $\beta_e$ for
the decay into final states with charged leptons becomes small. In this
case, the strict mass limits from Tevatron would not exclude the
existence of squarks in the mass range accessible to HERA. In fact,
searches at HERA have not found a signal and bounds on some of the
$\lambda'$ couplings have been derived from searches for the
characteristic lepton $+$ multijet final states which supersede previous
exclusion limits \cite{kuze}.

Most of the analyses done so far assume that only one of the couplings
$\lambda'_{ijk}$ is non-zero and only one squark state is in reach. A
more general scenario with two light squark states has been considered
in Ref.\ \cite{kon1} where it was shown that
$\tilde{t}_L$--$\tilde{t}_R$ mixing would lead to a broader $x$
distribution than expected for single-resonance production.  The
possibility of having more than one $\lambda'_{ijk} \ne 0$ was noticed
in Ref.\ \cite{belyaev} and deserves more theoretical study.


\section{Events with Isolated $\mu + p_{\rm T, miss}$}

$R_p$-violating supersymmetry has also played a role in the search for
explanations of the observation made by H1\footnote{No event of this
  type was observed by ZEUS \cite{zeusmuon}.} of five events with an
isolated $\mu$ and missing transverse momentum \cite{h1muon} (see also
\cite{meyer,WG3muon}). Events of this kind can originate from $W$
production followed by the decay $W \rightarrow \mu \nu_{\mu}$. Their
observed number is, however, larger than expected from the standard
model taking into account next-to-leading-order corrections to the
dominating resolved contribution from photoproduction \cite{spira}.
Moreover, their kinematic properties are atypical for $W$ production
\cite{meyer}.  An explanation in terms of anomalous $WW\gamma$ couplings
additionally has to face limits from Tevatron, LEP2 and ZEUS
\cite{zeusmuon} and leaves the question open why a similar excess of
events is not seen in $e + \rlap/p_T$ events.

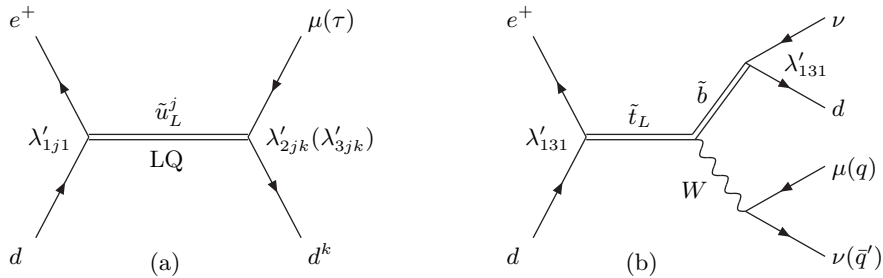
\begin{figure}[htb]
\small
\begin{picture}(135,100)(-10,0)
\ArrowLine(10,12)(30,50)
\ArrowLine(30,50)(10,88)
\ArrowLine(110,88)(90,50)
\ArrowLine(90,50)(110,12)
\Line(30,51)(90,51)
\Line(30,49)(90,49)
\put(0,92){$e^+$}
\put(0,2){$d$}
\put(113,92){$\mu(\tau)$}
\put(113,2){$d^k$}
\put(55,57){$\tilde{u}_L^j$}
\put(53,39){LQ}
\put(7,47){$\lambda'_{1j1}$}
\put(97,47){$\lambda'_{2jk}$($\lambda'_{3jk}$)}
\put(53,0){(a)}
\end{picture}
\begin{picture}(135,80)(-60,0)
\ArrowLine(10,12)(30,50)
\ArrowLine(30,50)(10,88)
\Line(30,51)(70,51)
\Line(30,49)(71,49)
\Line(70,51)(90,78)
\Line(71,49)(91.74,77)
\Photon(71,49)(90,22){2}{4}
\ArrowLine(120,95)(90,78)
\ArrowLine(90,78)(120,61)
\ArrowLine(120,39)(90,22)
\ArrowLine(90,22)(120,5)
\put(0,92){$e^+$}
\put(0,2){$d$}
\put(123,92){$\nu$}
\put(123,58){$d$}
\put(123,2){$\nu$($\bar{q}'$)}
\put(123,36){$\mu$($q$)}
\put(46,55){$\tilde{t}_L$}
\put(7,47){$\lambda'_{131}$}
\put(105,75){$\lambda'_{131}$}
\put(72,63){$\tilde{b}$}
\put(66,27){$W$}
\put(45,0){(b)}
\end{picture}
\caption[]{Possible decays of squarks produced in $e^+d$ scattering with 
  $R_p$-violating couplings leading to isolated $\mu$ + jet final
  states: (a) $\tilde{u}_L^j \rightarrow \mu d_k$ through
  $\lambda'_{2jk} \ne 0$; (b) $\tilde{t} \rightarrow \tilde{b} W$
  followed by $\tilde{b} \rightarrow \nu d$ via $\lambda'_{131} \ne 0$
  and $W \rightarrow \mu^+ \nu_{\mu}$ or $W \rightarrow$ 2 jets
  \protect\cite{kon2}.}
\label{figrp1}
\end{figure}

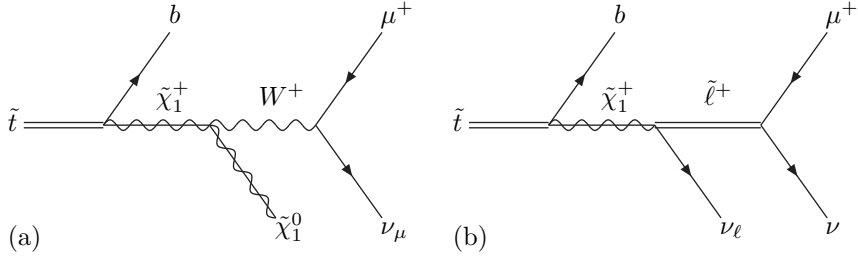
\begin{figure}[htb]
\begin{picture}(165,97)(-10,0)
\Line(10,49)(40,49)
\Line(10,51)(40,51)
\ArrowLine(40,50)(65,85)
\Photon(40,50)(80,50){2}{4}
\Line(40,50)(80,50)
\Photon(80,50)(105,15){2}{4}
\Line(80,50)(105,15)
\Photon(80,50)(120,50){2}{4}
\ArrowLine(145,85)(120,50)
\ArrowLine(120,50)(145,15)
\put(4,48){$\tilde{t}$}
\put(65,89){$b$}
\put(60,58){$\tilde{\chi}^+_1$}
\put(105,9){$\tilde{\chi}^0_1$}
\put(99,58){$W^+$}
\put(145,89){$\mu^+$}
\put(145,9){$\nu_{\mu}$}
\put(4,5){(a)}
\end{picture}
\begin{picture}(165,97)(-10,0)
\Line(10,49)(40,49)
\Line(10,51)(40,51)
\ArrowLine(40,50)(65,85)
\Photon(40,50)(80,50){2}{4}
\Line(40,50)(80,50)
\ArrowLine(80,50)(105,15)
\Line(80,49)(120,49)
\Line(80,51)(120,51)
\ArrowLine(145,85)(120,50)
\ArrowLine(120,50)(145,15)
\put(4,48){$\tilde{t}$}
\put(65,89){$b$}
\put(60,58){$\tilde{\chi}^+_1$}
\put(105,9){$\nu_{\ell}$}
\put(99,58){$\tilde{\ell}^+$}
\put(145,89){$\mu^+$}
\put(145,9){$\nu$}
\put(4,5){(b)}
\end{picture}
\caption[]{Possible decay chains of the stop leading to isolated muon +
  jet + missing $p_T$: $\tilde{t} \rightarrow b \tilde{\chi}_1^+$
  followed by (a) $\tilde{\chi}_1^+ \rightarrow \tilde{\chi}_1^0 \mu^+
  \nu_{\mu}$ \protect\cite{kon3}; (b) $\tilde{\chi}_1^+ \rightarrow
  \nu_{\ell} \mu^+ \nu$ \protect\cite{muon5}.}
\label{figrp2}
\end{figure}

The observation of $\mu+\rlap/p_T$ events could find an explanation in
$R_p$-violating scenarios if it is assumed that a stop, $\tilde{t}$, is
produced on-resonance at HERA.  Figures \ref{figrp1} and \ref{figrp2}
show examples for some of the possibilities. The process $ed \rightarrow
\tilde{t} \rightarrow \mu d^k$ (Fig.\ \ref{figrp1}a) which predicts
$\mu$ but no large $\rlap/p_T$ in the final state in gross disagreement
with the experimental observation, requires two different non-zero
$\lambda'$ couplings \cite{belyaev}. The relevant product
$\lambda'_{1j1} \lambda'_{2jk}$ would induce flavor changing neutral
currents and is therefore limited to small values for $1^{\rm st}$ and
$2^{\rm nd}$ generation quarks in the final state \cite{fcnclimits}. The
analogous process with a $\tau$ replacing the $\mu$ but followed by the
decay $\tau \rightarrow \mu \nu_{\tau} \bar{\nu}_{\mu}$ could also not
serve as an explanation since the decay-$\mu$ would be strongly boosted
in the direction of the $\tau$, i.e.\ the missing transverse momentum
would be correlated with the observed $\mu$ in contrast to the kinematic
properties of the H1 events. Moreover, hadronic decays of the $\tau$
would lead to an additional outstanding experimental signature and a
search for it at H1 was negative \cite{h1lq}.

The scenario shown in Fig.\ \ref{figrp1}b \cite{kon2} requires a
relatively light $b$ squark with $m_{\tilde{b}} \lsim 120$\,GeV, and
some fine-tuning in order to avoid too large effects on $\Delta \rho$ in
electroweak precision measurements. It could be identified by the
simultaneous presence of final states with $\rlap/p_T$ and multi-jets
from hadronic decays of the $W$.  Also the cascade decay shown in Fig.\ 
\ref{figrp2}a \cite{kon3} involving $R_p$-violation only for the
production of the $\tilde{t}$ resonance, not for its decay, seems
difficult to be achievable since it requires both a light chargino and a
long-lived neutralino. This, as well as the even more speculative
process shown in Fig.\ \ref{figrp2}b \cite{muon5} which requires
$R_p$-violation in the $L_i L_j E_k^c$ sector ($\lambda_{ijk} \ne 0$) as
well, can be checked from the event kinematics: assuming a value for the
mass of the decaying $\tilde{t}$, the recoil mass distribution must
cluster at a fixed value, the chargino mass. 

These speculations on a possible origin of the observed events within
$R_p$-violating supersymmetry are all linked to the presence of an
excess of events in NC scattering. The basic assumption is that a
squark, preferably a stop, is produced on-resonance; non-resonant stop
production would be too much suppressed.  Another type of explanation
not relying on this assumption was proposed in Ref.\ \cite{fritzsch}.
Events of the observed type could emerge after the production of a
single top quark followed by the decay chain $t \rightarrow bW$ and $W
\rightarrow \mu \nu$. The cross section of SM top production would be
much too small to explain the number of observed events, but the
presence of a coupling of the type of an anomalous-magnetic moment
inducing the transition $c \rightarrow t$ could enhance the cross
section considerably. However, the event rate would be still too small
unless a non-standard large $x$-behavior of the charm distribution
would be present, in addition. This scenario thus requires to open two
new fronts of non-standard physics. 


\section{Concluding remarks}

There are many scenarios of new physics for which deep inelastic
scattering experiments are most suitable to search for. Limits on
leptoquark of squark masses and their Yukawa or $R$-parity violating
couplings obtained at HERA will stay superior to those from other
experiments in many cases. 

The search for new physics effects relies in most cases on trustworthy
predictions from the standard model. In deep inelastic scattering this
includes the necessity to know parton distribution functions as
precisely as possible.  It is therefore a mandatory though nontrivial
task to combine the information from all available different experiments
in order not to run the risk of confusing modifications of parton
distributions with signs of new physics. With this in mind, the huge
amount of data expected from HERA experiments in the future is
guaranteed to play an indispensable role in the search for new physics
--- even in those cases where the most stringent limits are obtained at
other experiments.


\end{document}